\documentclass[sigconf]{acmart}
 
\usepackage{booktabs} 
\usepackage{subfigure}

\setcopyright{none}
\renewcommand\footnotetextcopyrightpermission[1]{}
\begin{document}
\title[Fashion Recommendation Using Relational Network]{Fashion Recommendation and Compatibility Prediction Using Relational Network}

\author{Maryam Moosaei}
\email{maryamoosaei@gmail.com}
\affiliation{%
  \institution{Visa Research}
  \city{Palo Alto}
  \state{California}
  \postcode{94306}
}

\author{Yusan Lin}
\email{yusalin@visa.com}
\affiliation{%
  \institution{Visa Research}
  \city{Palo Alto}
  \state{California}
  \postcode{94306}
}

\author{Hao Yang}
\email{haoyang@visa.com}
\affiliation{%
  \institution{Visa Research}
  \city{Palo Alto}
  \state{California}}
 

\begin{abstract}
Fashion is an inherently visual concept and computer vision and artificial intelligence (AI) are playing an increasingly important role in shaping the future of this domain. Many research has been done on recommending fashion products based on the learned user preferences. However, in addition to recommending single items, AI can also help users create stylish outfits from items they already have, or purchase additional items that go well with their current wardrobe. Compatibility is the key factor in creating stylish outfits from single items. Previous studies have mostly focused on modeling pair-wise compatibility. There are a few approaches that consider an entire outfit, but these approaches have limitations such as requiring rich semantic information, category labels, and fixed order of items. Thus, they fail to effectively determine compatibility when such information is not available. In this work, we adopt a Relation Network (RN) to develop new compatibility learning models, Fashion RN and FashionRN-VSE, that addresses the limitations of existing approaches. FashionRN learns the compatibility of an entire outfit, with an arbitrary number of items, in an arbitrary order. 
We evaluated our model using a large dataset of 49,740 outfits that we collected from Polyvore website. Quantitatively, our experimental results demonstrate state of the art performance compared with alternative methods in the literature in both compatibility prediction and fill-in-the-blank test. Qualitatively, we also show that the item embedding learned by FashionRN indicate the compatibility among fashion items.


\end{abstract}

%
%


%

\maketitle


\section{Introduction}
\label{sec:Introduction}

Fashion plays an important role in the society. People use fashion as a way of expressing individuality, style, culture, wealth, and status \cite{chao2009framework}. E-commerce fashion industry is expected to rise worldwide from \$481 billion USD revenue market in 2018 to \$712 billion USD by 2022\footnote{http://www.shopify.com/enterprise/ecommerce-fashion-industry, (accessed on 2018-07-17)}. This shows the increasing demands for online apparel shopping and motivates businesses to build more advanced recommendation systems. Many online retailers also started incorporating advanced recommendation systems to tackle the sophisticated fashion recommendation problem, such as StitchFix\footnote{http://www.stitchfix.com/}, asos\footnote{http://www.asos.com/} and Amazon Fashion\footnote{http://www.amazon.com/amazon-fashion}. This enormous e-commerce market has attracted researchers' attention in the artificial intelligence, computer vision, multimedia, and recommendation system communities \cite{mckinsey}.

Many research has been done using computational techniques to solve problems in fashion, e-commerce in particular. One most common line of research has been done on recommending single fashion items to consumers based on their purchase or browsing history. The most notable work is done by Kang et al, which they develop a neural network that learns users' preferences towards fashion products based on the visual information through Bayesian Personalized Ranking \cite{kang2017visually}.\footnote{http://wwd.com/business-news/business-features/jill-standish-think-tank-1202941433/}

\begin{figure}[t]
\centering
\includegraphics[width=0.45\textwidth]{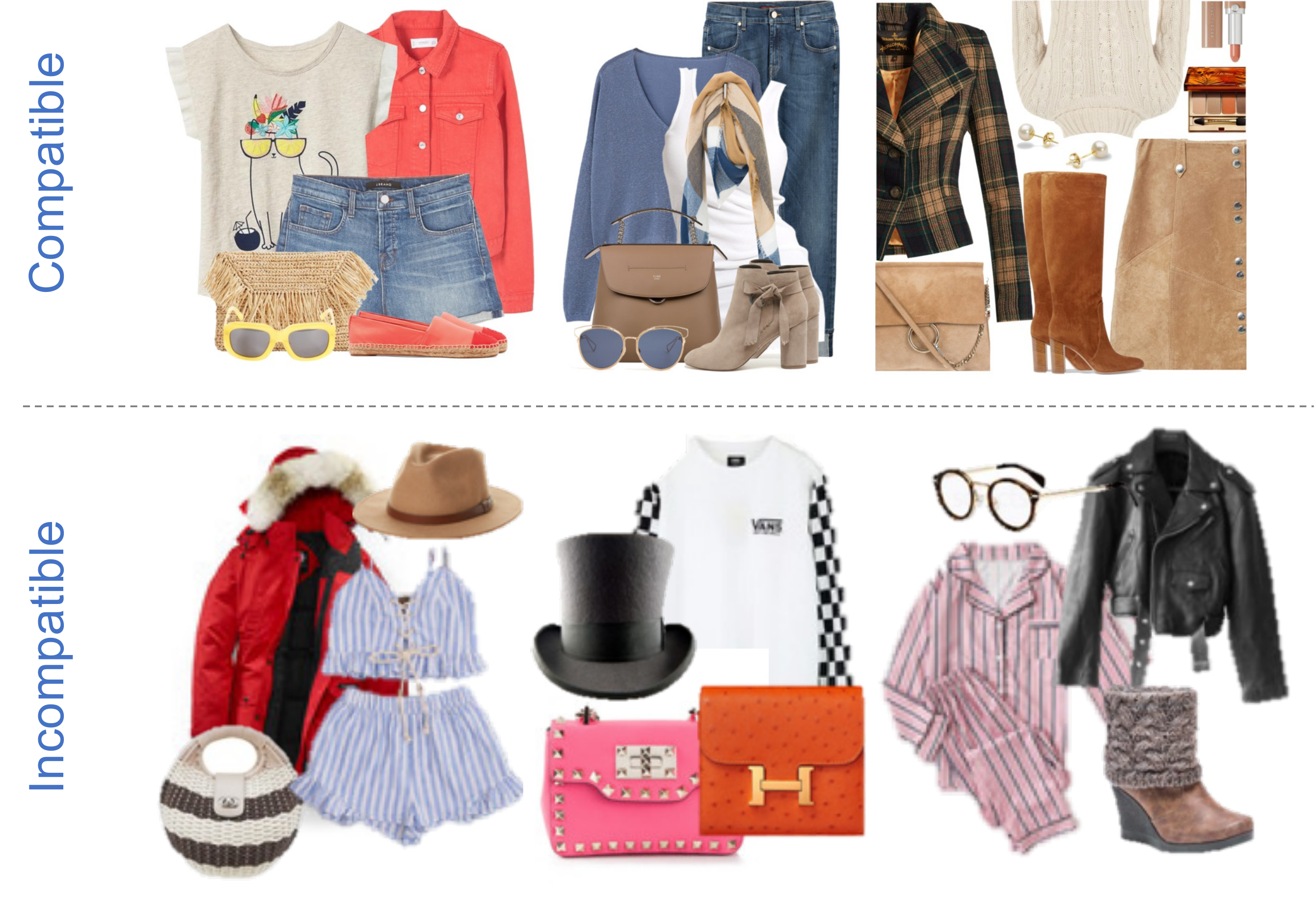}
\caption{Examples of compatible and incompatible outfits.}
\label{compatible}
\end{figure}

However, fashion recommendation is unique compare to other domains not only due to its heavily visual nature, but also because the concept of compatibility if more crucial than in any other types of products. People are often interested in purchasing items that match well together and compose a stylish outfit. Traditionally, fashion recommendation systems rely on co-purchased and co-clicked histories and recommend items based on similarity and user reviews. 
This requires going beyond retrieving similar items to developing a model that understands the notion of "compatibility" \cite{DBLP:conf/cikm/WanWLBM18}. Modeling compatibility is challenging because semantics that determine what is compatible and stylish are extremely complex and many factors such as color, cut, pattern, texture, style, culture, and personal taste play a role in what people perceive as compatible and fashionable. Developing an artificial intelligence algorithm that can learn the compatibility of items would considerably improve the quality of many fashion recommendation systems. Such systems will help customers decide what to wear every day, which alleviates the tedious task for non-fashion experts.


Nonetheless, recommending compatible items that form a fashion outfit includes several challenges. First of all, the item co-occurrence relationships are extremely sparse, hence a collaborative approach is hard based on such data nature. Leveraging the contents of items effectively is preferred. Secondly, compatibility is a very different concept from \emph{similarity}. Simply retrieving items that are similar to each other is merely enough to form fashion outfits. Many times, two items fit perfectly with each other in a fashion outfit, while when looking at them individually, they are visually extremely different. Thirdly, the number of items in fashion outfits varies. In most models, a fixed number of objects or fixed dimensionality is assumed as input. A model that encounters the different number of items as input is desired.

While visual features are commonly used in fashion recommendation, 
existing works on compatibility prediction have three main limitations:

\begin{enumerate}

\item Can only determine the compatibility of a pair of items and fail to work on outfits with an arbitrary number of items. Example methods with this limitation are \cite{vasileva2018learning,song2017neurostylist,veit2015learning}.

\item Need category labels (e.g., shirt, shoes) and rich attributes (e.g., floral, casual) in order to determine compatibility and will not work if such information is not available \cite{hsiao2017creating,li2017mining,han2017learning}.

\item Require a fixed order or fixed number of items to determine compatibility of an outfit. For example, Han et al. \cite{han2017learning} proposed a method for compatibility learning that requires items in all outfits to be carefully ordered from top to bottom and then accessories. 
\end{enumerate}

These limitations narrow the application of current methods. For example, many online retailers may lack detailed description or have noisy labels for fashion items. In addition, for items that are showcased at brick and mortar stores detailed descriptions are often not written on item tags.


To encounter the above challenges and limitations, in this paper, we use visual information of items to model fashion compatibility to optimize the content-based learning. We then, based on the concept of Relational Networks \cite{santoro2017simple}, build neural networks, FashionRN and FashionRN-VSE, that learn the relation between every pair of items, as well as take in different number of items as inputs. 
both FashionRN and FashionRN-VSE are to learn visual/textual relations between items of an outfit and use these relations to determine the compatibility of the outfits. The intuition behind using Relational Networks is that we can consider an outfit as a scene and the items of the outfit as objects in the scene. We are interested in learning a certain type of relation which is compatibility. 





To show the effectiveness of the proposed FashionRN and FashionRN-VSE, we evaluate our models on a collected Polyvore dataset, consisting of 49,740 unique fashion outfits. Through empirical experiments, we show that our proposed models perform well both quantitatively and qualitatively.

\underline{Quantitative results.} We design two evaluation tasks: fashion outfit compatibility prediction and fill-in-the-blank, and compare FashionRN and FashionRN-VSE's performances with other state-of-the-art models. We show that FashionRN-VSE achieves an Area Under Curve (AUC) of 88\% in the compatibility prediction task, compared to the second best comparing methods, Bi-LSTM with VSE, which achieves 72\%. Furthermore, FashionRN-VSE achieves an accuracy of 58\% in the fill-in-the-blank task, compared to the second best, SiameseNet, of 35\% in accuracy.

\underline{Qualitative results.} Besides learning the compatibility given an outfit, FashionRN and FashionRN-VSE can also generate item embedding from the hidden layer. Through visualization of the learned item embedding, we show that items that make sense to be put together in an outfit are closer to each other in the FashionRN embedding space. While comparing the visualization of the same items on embedding generated by current state-of-the-art CNN model, DenseNet, we see that DenseNet embedding place items that are visually similar (e.g., colors and shapes) but not necessarily compatible close to each other in the embedding space. This shows that embedding learned by FashionRN, besides the visual similarity, also captures the underlying compatibility.

 Our contributions are:


\begin{enumerate}
\item We developed FashionRN and FashionRN-VSE, a new line of compatibility learning framework based on Relational Networks \cite{santoro2017simple}. Our approach is independent of the number of items, order of items, and does not need semantic information and category labels.
\item We compared FashionRN and FashionRN-VSE to other state-of-the-arts in compatibility prediction and Fill In The Blank (FITB) task. We show that FashionRN outperforms the second best by 112.5\% and 148.5\% in the two tasks, repectively, while FashionRN-VSE outperforms the second best by 122.2\% and 165.7\%, respectively.
\item Through visualization, we find the item embedding learned by FashionRN well capture the underlying compatibility among fashion items, when compared to CNN models such as DenseNet that focus on the visual similarity.
\end{enumerate}

The reminder of this paper is organized as follows. Section \ref{relatedwork} reviews the related work. Section \ref{sec:methodology} describes our methodology. Section \ref{sec:experimental} presents our quantitative experimental results followed by our qualitative results in Section \ref{sec:qualitative}. We finally conclude this work in Section \ref{sec:conclusion}.

\section{Related Work}
\label{relatedwork}

In this section, we review the literature that are related to this work, which are fashion recommendation and relational networks.

\subsection{Fashion Recommendation}

There is a growing body of literature on fashion recommendation. Most of the available fashion recommendation systems use keyword search \cite{vaccaro2016elements}, purchased histories \cite{wang2011utilizing}, and user ratings \cite{kang2017visually,qian2014personalized} to recommend items. These methods do not consider visual appearance of items which is a key feature in fashion. 

To address this limitation, several research groups have worked on incorporating visual information in fashion recommendation systems, mainly with the purpose of recommending similar items to an image query \cite{jing2015visual, he2016fashionista,chao2009framework,tautkute2018deepstyle,DBLP:conf/iccv/RenSLMF17}, and recommending aesthetics based on personal preferences \cite{DBLP:conf/cikm/DengCFNY17,DBLP:conf/cikm/SkopalPKGL17}. Similarity based fashion recommendation systems are useful for finding substitutes for an item (e.g., finding a shirt with the same style but different brand or price) or matching street images to online products \cite{chao2009framework,hadi2015buy}. However, many times users are interested in searching for different category of items which are compatible and are in harmony. This requires going beyond similarity based methods and modeling more complex concepts such as compatibility and aesthetics. 

Many humans are expert in detecting whether an outfit looks compatible or something is "off" by simply looking at its appearance. For example, even though compatibility is a subjective concept, most people would agree that the outfits shown in Figure \ref{compatible} are all well composed and stylish. Research has shown that computer vision and artificial intelligence algorithms are also able to some extent learn the notion of compatibility \cite{song2017neurostylist,veit2015learning,han2017learning,he2018fashionnet}. For example, Iwata et al. used a topic model to find matching tops (e.g., shirt) for bottoms (e.g., jeans) using a small human annotated dataset collected from magazines \cite{iwata2011fashion}. 

Veit et al. \cite{veit2015learning} used images of co-purchased items from an Amazon dataset 
to train a Siamese neural network \cite{hadsell2006dimensionality} for predicting compatibility between pairs of items. Song et al. showed that integrating visual and contextual information can improve compatibility prediction \cite{song2017neurostylist}. To exploit the pair-wise compatibility between tops and bottoms they learned a latent compatibility space by employing a dual autoencoder network \cite{ngiam2011multimodal} and a Bayesian Personalized Ranking (BPR) framework \cite{rendle2009bpr}. 

Lin et al.  developed a model that is not only capable of matching tops with bottoms, but also is able to generate a sentence for each recommendation to explain why they match \cite{lin2018explainable}. Instead of a dual auto-encoder network, they used a mutual attention mechanism to model compatibility and a cross-modality attention module to learn the transformation between the visual and textual space for generating a sentence as a comment. 

Vasileva et al. \cite{vasileva2018learning} extended state-of-the-art in compatibility learning by answering novel queries such as finding a set of tops that can substitute a particular top in an outfit (high compatibility), while they are very different (low similarity). To do this, they jointly learned two embedding spaces, one for item similarity and the other for item compatibility.

All of the aforementioned methods are pair-wise and focus on learning compatibility between "tops" and "bottoms". These methods fail to consider an entire outfit with an arbitrary number of items. To address this limitation, Han et al. \cite{han2017learning} and Jiang et al. \cite{jiang2018ask} considered an outfit as a sequence (from top to bottom and then accessories) and each item in the outfit as a time step \cite{han2017learning}. They trained a bidirectional LSTM (Bi-LSTM) model to sequentially predict the next item conditioned on previous ones, learning their compatibility. They used attribute and category information as a regularization for training their model.
Treating outfits as a sequence and using an LSTM-based model does not respect the fact that sets are order invariant. Consequently, it requires carefully sorting of items in all outfits in a consistent order based on their category labels. Otherwise, a compatible top-bottom may be detected as incompatible if one changes their order to bottom-top.

Li et al. developed a model that considers outfits as order-less sets. Given a collection of fashion items, their method can predict popularity of a set by incorporating images, titles, and category labels \cite{li2017mining}.


In a recent work, Hsiao and Grauman \cite{hsiao2017creating} proposed an unsupervised compatibility learning framework which uses textual attributes of items. The researchers employed a Correlated Topic Model (CTM) \cite{blei2006correlated} from text analysis to learn compatibility. They considered an outfit as a document, visual attributes (e.g., floral, chiffon) as words, and style as a topic. Their model learns the composition of attributes that characterizes a style. For example, a formal blazer is more likely to be combined with a pair of jeans than a floral legging.

While a fair number of studies are available on compatibility prediction, existing methods are mostly pair-wise and a few studies which consider an entire outfit \cite{hsiao2017creating,han2017learning}, are either not order invariant with respects to the items in an outfit \cite{han2017learning}, or require rich contextual data including explicit category labels, whether extracted from item descriptions or human annotated \cite{hsiao2017creating}. Hence, in our work, we explored a new visual compatibility learning framework that would consider an entire outfit with an arbitrary number of items with an arbitrary order. Our model can work without category labels or semantic attributes.

\begin{figure*}[t]
\centering
\includegraphics[width=0.9\textwidth]{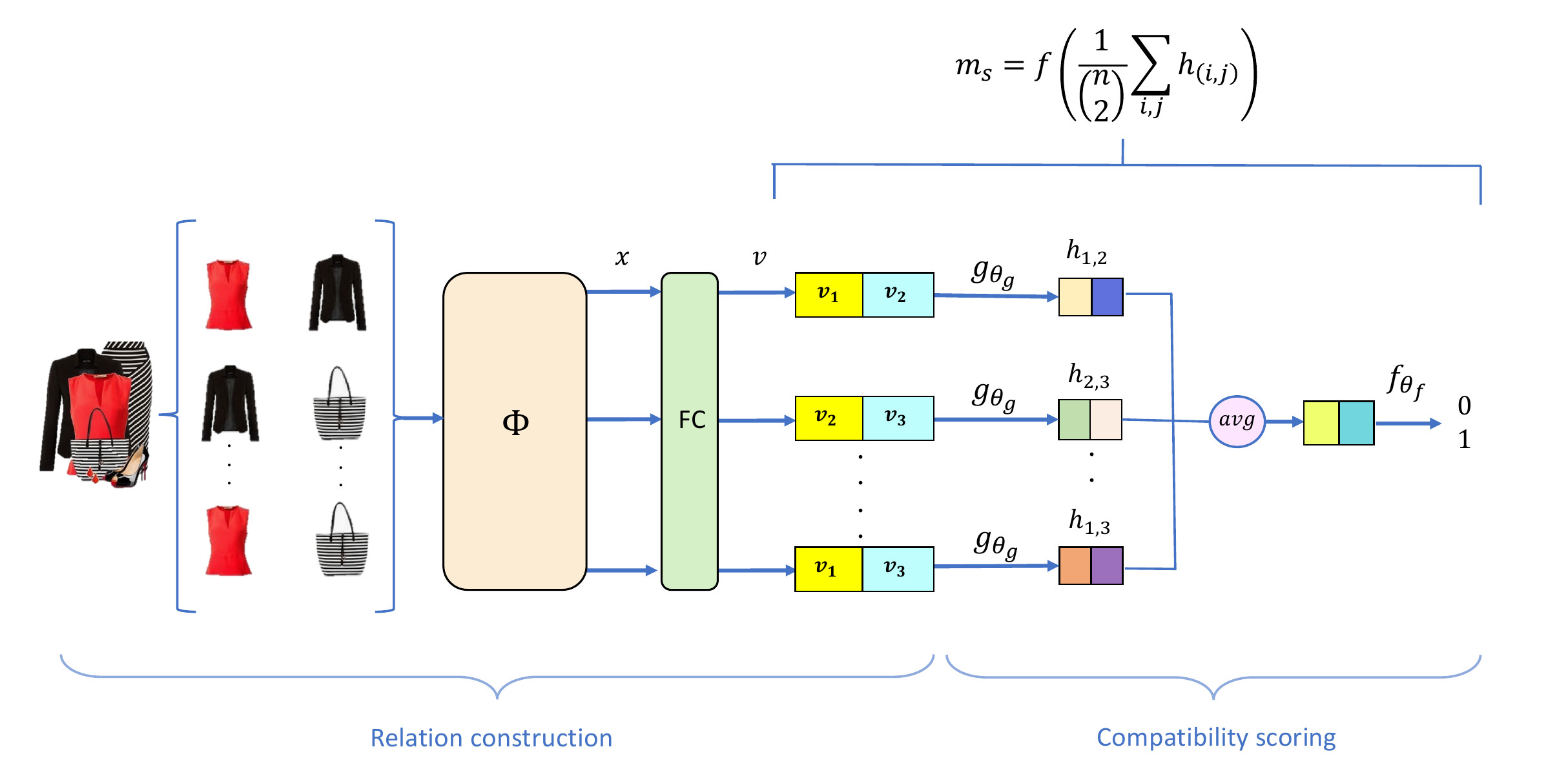}
\caption{Model design of FashionRN.}
\label{diagram}
\end{figure*}

\subsection{Relational Networks}


Many factors such as style, texture, material, and color contribute to compatibility and the relation between these factors is non-linear. In this work, we develop Fashion RN by modifying a Relational Network (RN) to learn a non-linear space that can predict the compatibility of an outfit. 

Previous findings suggest that relational reasoning is "baked" into RNs, similar to learning sequential dependencies which is built in recurrent neural networks \cite{santoro2017simple}. Different variations of RNs have been successfully applied to answering semantic based questions about dynamic physical systems. For example, Santoro et al. modified an RN architecture and showed that given an image of a scene, an RN combined with an LSTM can answer relational questions such as "\textit{Are there any rubber things with the same size of the yellow cylinder?}"

The input to an RN is a set of objects, but the definition of an object is flexible and not specified. For example, Santoro et al. used a CNN network to convert images of physical systems into $k$ feature maps of size $d*d$ \cite{santoro2017simple}. They then considered each row of the feature map as an object. Therefore, in their work, an object could be a part of the background, a particular physical object, or a texture. The object-object relation in their work was question dependent. Thus, their RN architecture was conditioned on a question embedded by an LSTM. Each pair of objects was concatenated with the question embedding before going into the RN.

The intuition behind our approach is that humans do not need to know the textual description of items in an outfit (see Figure \ref{compatible}) and their category labels in order to know if it looks compatible. Humans can detect compatibility in a visual scene by looking at it. In fact, many of the textual attributes (e.g., floral, shirt, casual) can be implicitly learned from visual information. Moreover, sets are order invariant. For example, humans do not need to see the items of an outfit in a specific order (e.g., always seeing pants before seeing shirts) in order to detect their compatibility.  Therefore, in this work we try to model similar intelligence by developing a compatibility learning method that is based on visual information and does not require labeling clothing attributes or feeding items in a fixed order.

Our network, Fashion RN, is based on Relational Networks (RNs) which are architected for relational reasoning \cite{raposo2017discovering}. Santoro et al. successfully applied RNs to text-based question answering about scenes \cite{santoro2017simple}. We considered compatibility as a particular type of relation and explored developing an RN inspired architecture that can learn the compatibility between items in an outfit.


\section{Compatibility Learning with Relational Network}
\label{sec:methodology}

In this section, we propose our model, FashionRN, that learns the compatibility among fashion items in a fashion outfit. We also propose its variant, FashionRN-VSE. For the ease of understanding, we summarize the symbols used in this paper in Table \ref{table:symbol_definition}.

\begin{table}[htb!]
\caption{Symbol definition.}
\label{table:symbol_definition}
\begin{tabular}{@{}cl@{}}
\toprule
\textbf{Symbol} & \textbf{Definition}              \\ \midrule
$\mathcal{D}$               & Dataset                          \\
$\mathcal{I}$               & Item set                         \\
$\Phi$             & CNN model                        \\
$x$               & High-dimensional visual features \\
$v$               & Low-dimensional visual features  \\
$d$               & Textual embedding                \\
$h$               & Relation embedding               \\
$f, g$            & Fully-connected layers           \\ \bottomrule
\end{tabular}
\end{table}


\subsection{Problem Formulation and Model Intuition}

We assume the compatibility of a fashion outfit to be based on the relation among all of the items included in an outfit. To learn the compatibility of fashion outfits, we formulate our problem into a binary classification problem. Let $S = \lbrace i_1,i_2,…i_n \rbrace$ be a fashion outfit, where each $i \in \mathcal{I}$ is an item in this set. The dataset $\mathcal{D} = \lbrace S \rbrace$. Given an $S$, predict whether it is a compatible fashion outfit or not.

The learning of fashion outfit compatibility can be thought of as follows. For a fashion outfit, we measure the compatibility of each pair of items in the outfit, and eventually aggregate all of the pairs' compatibility scores to obtain the overall outfit compatibility score. To achieve this, we propose two models: FashionRN and FashionRN-VSE, which we describe in detail in the following.

\begin{figure*}[t]
\centering
\includegraphics[width=0.9\textwidth]{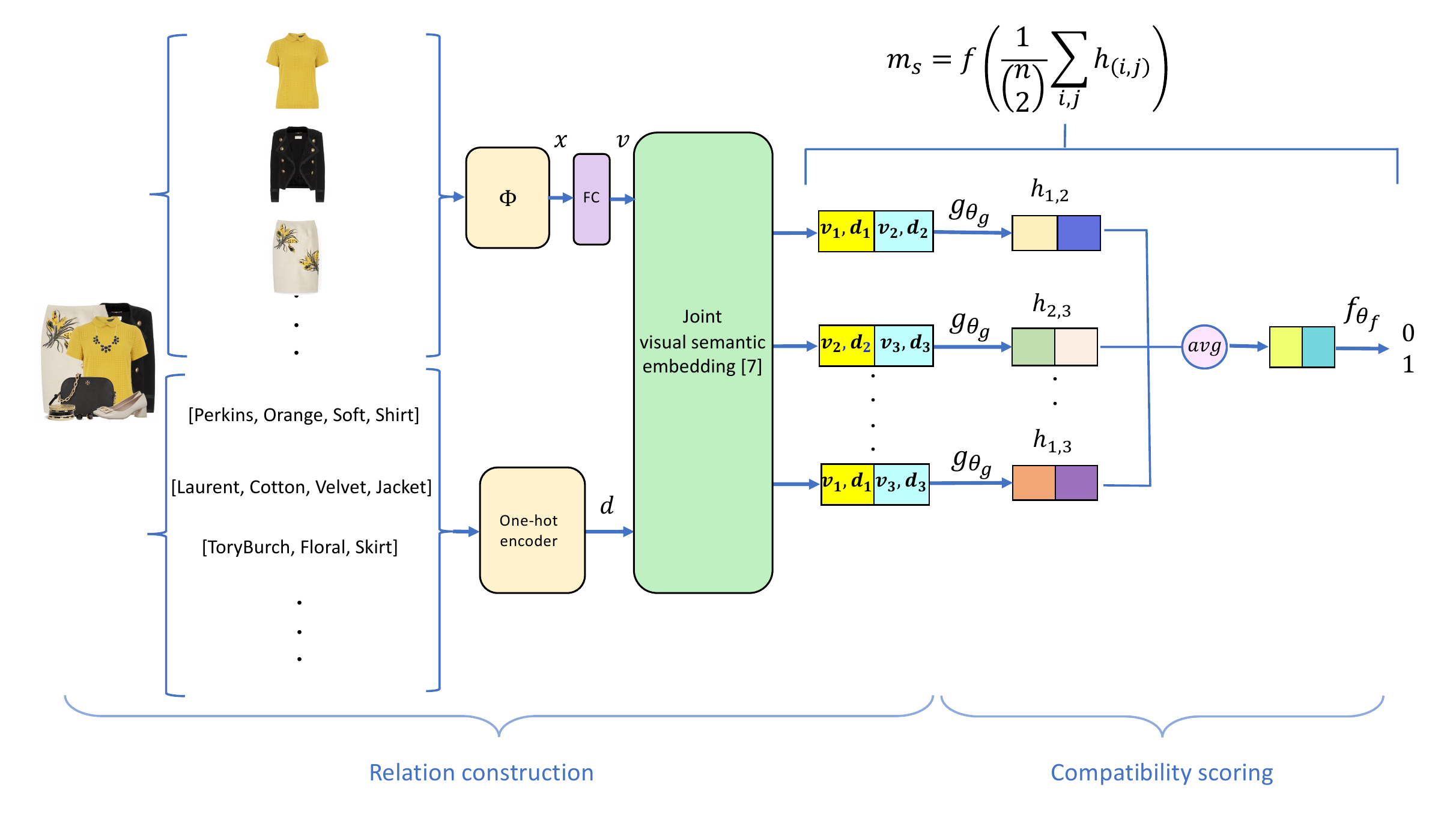}
\caption{Model design of FashionRN-VSE.}
\label{joint}
\end{figure*}


\subsection{FashionRN}
\label{sec:compatibility learning}

We design FashionRN based on the concept of the relational network architecture. In our model, an outfit is treated like a scene and its items are treated like the objects in the scene. Therefore, as opposed to Santoro et al. who consider rows of a CNN feature map, extracted from the entire scene, as objects; we consider images of items in an outfit as objects and use a DenseNet to transform them into feature vectors. Additionally, we are interested in learning one specific type of visual relation, compatibility, which is not question dependent and therefore we do not need any LSTM model. 

Our FashionRN consists of two parts, as shown in Figure \ref{diagram}. The first part, \emph{relation construction}, learns the non-linear relation between each pair of items in an outfit and the second part \emph{compatibility scoring}, combines all the pair-wise relations to learn the compatibility of the entire outfit. 

\subsubsection{Relation Construction}



First, the images of items are passed through a pre-trained CNN model of choice $\Phi$ (e.g., DenseNet) to produce high-dimensional visual feature vectors, $\mathbf{x}$.
$\mathbf{x}$ is then passed through a fully connected (FC) layer, which serves two purposes. It down sizes the feature vectors and learns a latent space from dimensions that correspond to fashion styles and contribute to compatibility. The reduced-dimensional features are denoted as $v$.


After generating the lower-dimensional visual features $v$, the relation between each pair of items in $S$ is constructed as follows. For each pair of items $(i,j) \in S$, we concatenate their visual features and passed through a FC layer $g$ to generate relation embedding $h$.

\begin{align}
    h_{(i,j)} = g([ v_i || v_j ])
\end{align}

\subsubsection{Compatibility Scoring}

After the relation construction, we model the compatibility among all the pairs of items in $S$ as follows.

\begin{align}\label{eq:compatibility_score}
 m_s = f \Big( \frac{1}{\binom{n}{2}}\sum_{i,j}h_{(i,j)}  \Big) 
\end{align}

\noindent where $m_s$ is the compatibility score of outfit $S$. Both $f$ and $g$ are based on multiple non-linear functions with parameters $\theta_f$ and $\theta_g$. In our work, $ f_{\theta_f} $ and $ g_{\theta_g} $ are multi-layer perceptrons (MLPs) and we want to learn the parameters $\theta = \lbrace \theta_f, \theta_g \rbrace$ such that they can predict the compatibility between fashion items. The output of $g_\theta$ is the "relation" \cite{santoro2017simple}. Thus, $g_\theta$ learns the relation between the visual appearances of $v_i$ and $v_j$. 

\subsection{FashionRN-VSE}

While some studies learn compatibility using visual information \cite {vasileva2018learning,veit2015learning}, others have suggested combining textual data with visual data can improve the performance of compatibility prediction  \cite{han2017learning,li2017mining,hsiao2017creating,song2017neurostylist}. We hence propose a variant of FashionRN, which combines the concept of Visual Semantic Embedding (VSE) proposed by Han et al. \cite{han2017learning} We name this model FashionRN-VSE.

The diagram of this method is presented in Figure \ref{joint}. VSE produces image embedding ($v_i$) and description embedding ($d_i$) for an item $i$. $v_i$ is produced by passing through a CNN model of choice $\Phi$ as in FashionRN, while $d_i$ is produced by encoding each word in the outfit description to a one-hot encoding. $v_i$ and $d_i$ for each item in an outfit are concatenated and fed into FashionRN-VSE. The compatibility stays the same as Eq. (\ref{eq:compatibility_score}), while the relation embedding for FashionRN-VSE is reformulated as follows.
%

\begin{align}
    h_{(i,j)} = g \big( (v_i || d_i) || (v_j || d_j) \big)
\end{align}

With the consideration of textual information, FashionRN-VSE not only considers the visuals of fashion items, but also more detail information beyond what can be observed from the images. These information include: brands, texture, material, and even price point, etc. We believe through capturing these information, FashionRN-VSE can better learn the compatibility of fashion items in a fashion outfit.



\subsection{Design Options and Time Complexity}

Our proposed models enable various design and usage options. First of all, depends on one's data richeness, one can choose to use FashionRN if only visual information is available, and FashionRN-VSE is both visual and textual information are available. 

Secondly, RNs are order invariant. Therefore, it does not matter in which order the outfit items are passed to the network. Although to detect the compatibility of an outfit we consider the relation between all of its items, using RNs gives the flexibility to consider only some of the item pairs, or to put greater weights on some of them. For example, if an item (e.g., a handbag) is the center piece of an outfit and one likes to compose an outfit that magnifies this piece, we can put greater weights on the relations that involve this item. 

Besides the flexibility, our proposal is also efficient time-complexity-wise. Our compatibility learning framework can be applied to outfits with an arbitrary number of items. The time complexity of calculating the compatibility of an outfit with $n$ items is $O(\binom{n}{2})$ with respects to the number of the items in the outfit\footnote{We empirically found that the order of objects in each pair does not impact the accuracy and thus our time complexity is  $O(\binom{n}{2})$ and not $O(n^2)$}. However, considering that outfits have limited number of items (less than 12 in our dataset), this time complexity will remain linear $O(n)$. 
Also, developing a compatibility framework based on RNs eliminates the need of passing item category labels as input to the network, as the network itself is able to implicitly learn such information. 

\subsection{Parameter Learning}
\label{sec:Implementation}

The parameters $\theta_g$ and $\theta_f$ are learned through back propagation using a cross-entropy loss function as follows:

\begin{align}
\label{eq:30}
\mathcal{L} (\theta_g, \theta_f) = -\sum_{i}^{|\mathcal{B}|} (y_i \log(p_i)+(1-y_i) \log(1-p_i))
\end{align}
\noindent where $\mathcal{B}$ is one batch of training, $y_i$ is the ground truth label and $p_i$ is the predicted label ($m_s$) of the $i$\textsuperscript{th} outfit.

To learn the parameters, all of the outfits in the datasets are viewed as positive samples, with $y$ expected to be 1s. To create negative samples, we randomly select numbers of items to create artificial outfits, and set their labels $y$ to be 0s.

\section{Evaluation}
\label{sec:experimental}

To examine the effectiveness of FashionRN and FashionRN-VSE, in this section, we empirically test their performances on two prediction tasks: compatibility prediction and fill-in-the-blank test, on a large fashion outfit dataset, and compare with other state-of-the-arts.



\subsection{Dataset}
\label{sec:Dataset}

Learning compatibility of fashion outfits requires a rich source of data which can be collected from online fashion communities such as Polyvore, Chictopia\footnote{http://www.chictopia.com}, and Shoplook\footnote{https://www.shoplook.io}. On these websites, users can create stylish outfits and look at million of outfits created by others. Such rich fashion data can be used to train 
neural networks to learn different fashion concepts and automatically create stylish outfits. Polyvore is a great source of data especially for our work because it has images of items with clear background and descriptions.

Researchers have used data from Polyvore for various studies \cite{vaccaro2016elements,li2017mining,lee2017style2vec,vasileva2018learning}. However, some of their datasets are not open source (e.g.,\cite{li2017mining}) or have a small size (e.g., \cite{han2017learning,song2017neurostylist}). Thus, we collected our own dataset from Polyvore. To ensure the quality, we collected outfits from users who are highly popular on Polyvore and have at least 100K followers. For each item we saved a 150 x 150 image and item description.
We cleaned the dataset by excluding items that are not clothing (e.g., furniture) using their metadata. Then, we removed any outfit that is left with only one item. The remaining dataset had 49,740 outfits and 256,004 items. The collected outfits have arbitrary number of items ranging from 2 to 12, but on average each outfit has five items. We used 70\% of our data for training (34,818 sets), 15\% for validation (7,461 sets) and 15\% for testing (7,461 sets).

\underline{Negative Sample Creation.} The data collected from Polyvore includes compatible outfits (positive class). 
Following the methodology of \cite{han2017learning} we created our negative class by randomly picking items from different outfits. While, these outfits are not guaranteed to be incompatible, they have a lower probability of compatibility compared to outfits that have been created by fashion experts on Polyvore and therefore our network should assign lower compatibility scores to these randomly composed outfits. We created an incompatible outfit per each positive outfit. This resulted in overall 69,636 sets for training (positive and negative), 14,922 sets for validation and 14,922 sets for testing.


\subsection{Experiment Setting}

We choose DenseNet as our CNN model $\Phi$ since at the time of writing, it is the state-of-the-art. DenseNet generates image features $x$ of dimension 94,080. We design the FC layer $f$ to output 1000-dimensional features, so that $v\in R^{1000}$. 

In our work, $f$ and $g$ are both multi-layer perceptrons (MLPs). $g$ has four layers with size 512, 512, 256, 256 and $f$ has three layers with size 128, 128, 32. Therefore, $\Theta _g \in R^{2000*256}$ and $\Theta _f \in R^{256*32}$. At the end we used a softmax layer for classification. We used layer normalization and ReLU activation for all the layers of $f$ and $g$. We set dropout rate to 0.35 for all the layers except the last layer of $f$.

We set the learning rate to 0.001 and the batch size to 64. Therefore, each mini batch included 64 fashion sets. Finally, we trained our model until the validation loss stabilized which took 19 epochs. Our model is implemented using Tensorflow, and Adam optimizer is used to learn the parameters. All our experiments are run on GPU Tesla P100-PCIE-16GB.


\subsection{Prediction Tasks}

An effective compatibility model, given an unseen fashion outfit, should accurately score the outfit based on how the items included match with each other. Besides, given an incomplete fashion outfit, it should also be able to provide suggestion on fashion item to fill in. With such objectives in mind, we design two prediction tasks to evaluate the effectiveness of FashionRN and FashionRN-VSE.

We evaluated our method using the large dataset we collected from Polyvore (Section \ref{sec:Dataset}). We performed two tests: 
\begin{itemize}
\item \textbf{Compatibility prediction test}: predict the compatibility score of a given fashion outfit. This test is a binary classification task, where the model should answer true if the given outfit is compatible, and false otherwise.
\item \textbf{Fill in the blank (FITB) test}: given an outfit and a number of candidate items, find the item that matches best with the existing items in the outfit. This test is a retrieval task, where given an incomplete fashion outfit, and a list of candidate fashion items, the model aims to score all of the candidate items and return the item with the highest compatibility score with the incomplete fashion outfit.
\end{itemize}
These two tests are commonly used in the fashion recommendation literature for evaluating compatibility learning methods \cite{han2017learning,hsiao2017creating,vasileva2018learning}. 

\subsection{Comparing Methods}

To demonstrate the effectiveness of our proposed method, we compared our results with the following approaches and demonstrate our results in Table \ref{table1} and Table \ref{table2}. 
We evaluated these methods on the dataset described in Section \ref{sec:Dataset}. For each method, we used the authors' codes and their reported set of parameters. We have considered compatibility prediction as a binary classification task and have calculated Area Under Curve (AUC) score to compare these methods.

\begin{itemize}

\item\textbf{Bi-LSTM + VSE} \cite{han2017learning}: A fashion outfit is considered as a sequence from top to bottom and a Bi-LSTM model is jointly trained with a visual-semantic embedding (VSE) model to learn compatibility.




\item \textbf{SiameseNet} \cite{veit2015learning}: SiameseNet uses a Siamese CNN to transform images into an embedding space in which compatible items are close to each other and are far away from incompatible items. After training the network uses a contrastive loss, the distance between item embeddings is used for estimating their compatibility. To compare with this network, we created compatible pairs by selecting items from the same outfit. Incompatible pairs were created by selecting items from different outfits. To measure the compatibility of an outfit using SiameseNet, we averaged the compatibility scores of all of the item pairs in that outfit.


\item \textbf{BPR-DAE} \cite{song2017neurostylist}:  A latent compatibility space is learned by employing a dual autoencoder (DAE) network and a Bayesian Personalized Ranking (BPR) framework. We trained BPR-DAE similar to SiameseNet and considered the average compatibility score of all the item pairs in an outfit as its compatibility score. 


\item \textbf{RAW-V}: The compatibility score of an outfit $S$ is measured based on the raw visual features of its items as:

\begin{align}
\label{eq4}
m_s = \frac{1}{\binom{n}{2}}\sum_{i,j} d(v_i, v_j)
\end{align}

$v_i$ and $v_j$ are the visual feature representations of items $i$ and $j$, extracted from a fined-tuned DenseNet \cite{huang2017densely} and $d(v_i , v_j) = v_i \cdot v_j$ is the cosine similarity between items $i$ and $j$. The compatibility of an outfit is obtained by averaging pair-wise compatibilities of all the pairs in the outfit. 


\item \textbf{VSE}: We learned the joint visual semantic embedding proposed by Han et al. \cite{han2017learning} and measured compatibility similar to RAW-V.



\item \textbf{FashionRN}: Our proposed model considers a fashion outfit as a scene, and items in the outfit as objects in the scene. It then learns the compatibility of an outfit with arbitrary number of items using a Relational Network. 

\item \textbf{FashionRN-VSE}: Our proposed model that builds on top of FashionRN, and adds in the component of VSE.




\end{itemize}

The first four methods are popular in the literature for learning compatibility and the rest are for understanding how different components contribute to compatibility. As our method is mainly based on visual information, we did not compare our method with approaches which only rely on semantic information for learning compatibility \cite{hsiao2017creating}. 



\begin{figure}[t]
\centering
\includegraphics[width=0.9\linewidth]{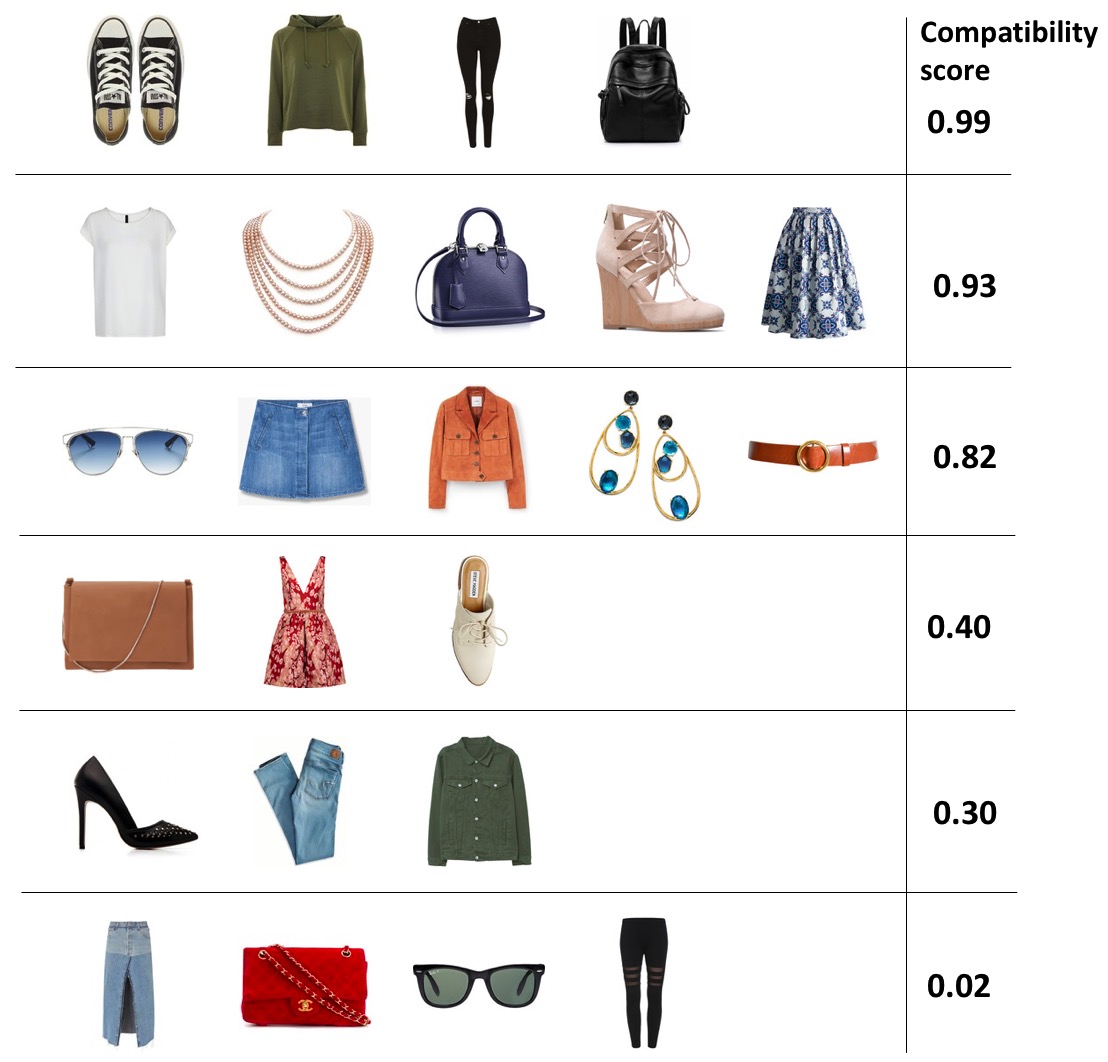}
\caption{Example test outfits in our compatibility prediction task and their scores.}
\label{compat}
\end{figure}

\subsection{Compatibility Prediction}
\label{sec:compatibility prediction}

In this task, a number of items are given as input and we aim to find their compatibility score. For items that are compatible with each other, the model should answer yes, and false otherwise. This enables a recommendation system to recommend items based on their compatibility with a query or with items in a shopping cart. In addition, users can create their own outfits and know their compatibility. 

\begin{table}[htb!]

\caption{Performance of different approaches on the compatibility prediction test.} 
\label{table1}
\centering

{\begin{tabular}{ l  | c }
\hline \hline
Approaches &  AUC  \\ 
\hline 
Bi-LSTM + VSE \cite{han2017learning} & 0.72  \\
SiameseNet \cite{veit2015learning} & 0.48 \\
BPR-DAE \cite{song2017neurostylist} & 0.53\\
RAW-V & 0.61 \\
VSE & 0.45 \\
Fashion RN  & \textbf{0.81} \\
Fashion RN + VSE   & \textbf{0.88}  \\

\hline
\end{tabular}  }  
\end{table}

Table \ref{table1} shows the performance comparison among different approaches for compatibility prediction task. This table shows that both of our models, FashionRN and FashionRN-VSE, achieve the best performance among the comparing methods, 
including the Bi-LSTM method which requires both visual and semantic information. This is because our Relational Network based model is inherently able to learn a variety of relations between items including their categories without requiring to have access to explicit semantic attributes and category labels. This is specially useful when semantic information is not available or is very noisy. 


We also observe that Bi-LSTM performance decreases on our dataset. This is likely due to our test dataset size (14,922 outfits) which is much larger than the test dataset (3,076 outfits) used by the authors \cite{han2017learning}.

Table \ref{table1} shows that our method performs better than the two comparing pair-wise methods (SiameseNet and BPR-DAE). This finding suggests that pair-wise methods fail to work well on learning the compatibility of outfits with more than two items. This is because a linear combination of pair-wise compatibility scores (e.g., averaging all the pair-wise scores) fails to capture the compatibility of an entire outfit. In our work, although we start by learning the relation between item pairs, we combine the pair-wise relations and pass them through multiple nonlinear layers to learn more powerful feature representations from an entire outfit. This can determine the compatibility of an outfit more accurately than simply averaging all the pair-wise compatibility scores.





\begin{table}[t]

\caption{Performance of different approaches on FITB test.} 
\label{table2}
\centering

{\begin{tabular}{ l  | c   }
\hline\hline
Methods & Accuracy\\ 
\hline
Bi-LSTM + VSE \cite{han2017learning} & 0.34  \\
SiameseNet \cite{veit2015learning} & 0.35 \\
BPR-DAE \cite{song2017neurostylist} & 0.20\\
RAW-V & 0.35 \\
VSE & 0.33 \\
Fashion RN  & \textbf{0.52} \\
Fashion RN + VSE   & \textbf{0.58}   \\

\hline
\end{tabular}}    
\end{table}

Figure \ref{compat} shows qualitative results of our model for compatibility prediction. Compatible outfits have two or more non redundant items that have well-matching colors and share similar style. From Figure \ref{compat}, we can observe that our method can effectively predict if a set of items make a compatible outfit.  For example, items in the first row, are all black/green and share a casual/sportive style and therefore they have a high compatibility score; Items in the second row have chic/formal style and are all cream or dark blue which together create a stylish contrast and therefore have a high compatibility score.

Items of incompatible outfits may have inconsistent styles or colors. Incompatible outfits may also have redundant items such as two shirts. We can observe that our method is able to capture such concepts from visual information. In contrast to Bi-LSTM model, we do not need to feed any category labels or attributes (e.g., men, women, shirt, shoes), to our model to explicitly teach it that for example a men's shirt is incompatible with a woman's skirt, or an outfit with redundant items is incompatible. Our model is able to implicitly learn such information. For example, items in the fourth row do not have compatible colors/patterns and therefore have received a low compatibility score; items in the fifth row have compatible colors, but a man's shirt and a pair of men's jeans do not match with women's heels. Thus, this outfit has also received a low score; finally, the outfit in the last row has two bottoms (skirt and leggings) and our network has given a low compatibility score to this outfit.



\begin{figure}[t]
\centering
\includegraphics[width=0.9\linewidth]{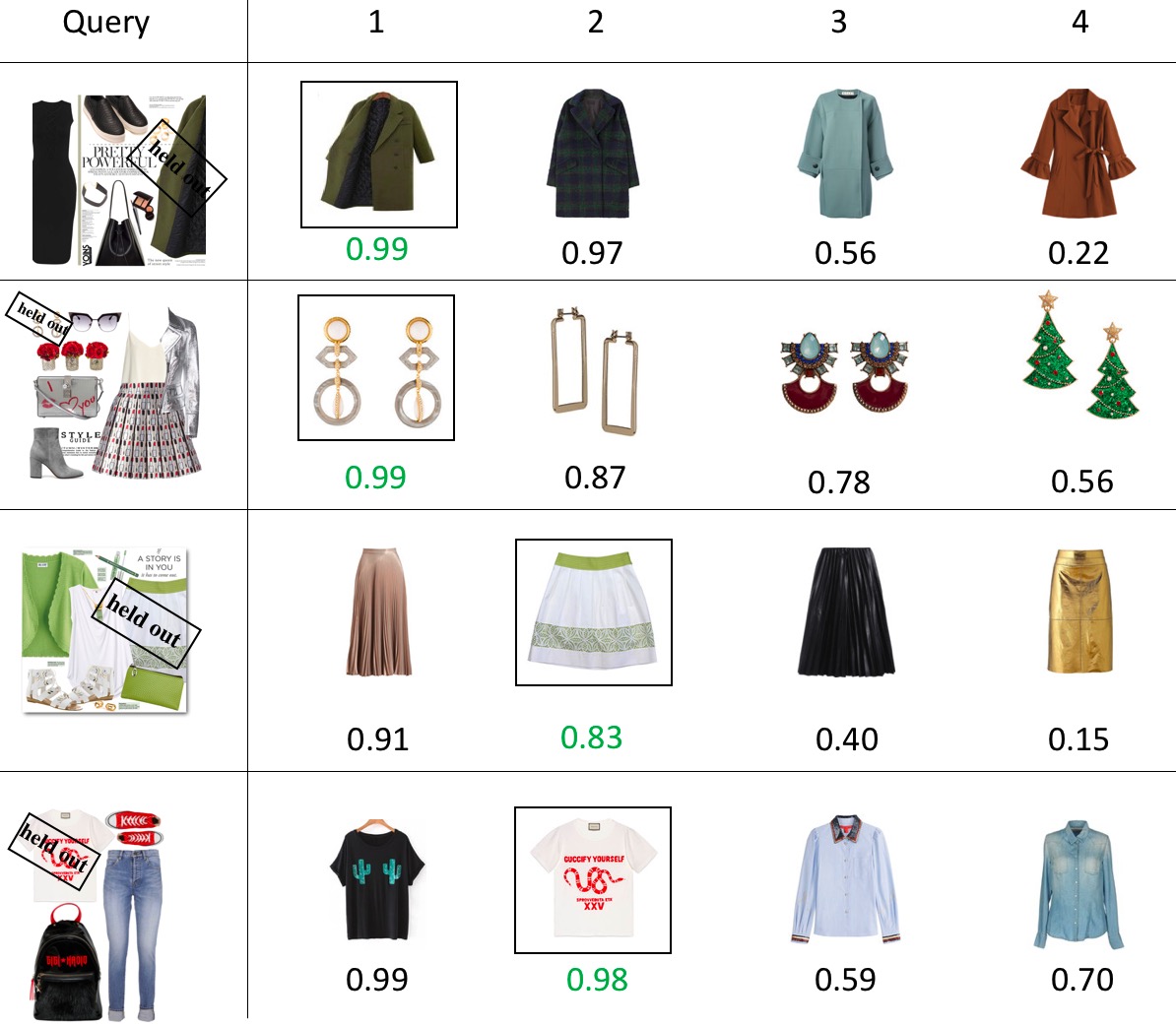}
\caption{Example results from the FITB task using Fashion RN model. Items in each row are ranked based on their output scores and held out items are highlighted in rectangles.}
\label{fitb}
\end{figure}

\begin{figure*}[t!]
    \centering
    \subfigure[DenseNet]{
        \includegraphics[width=.9\linewidth]{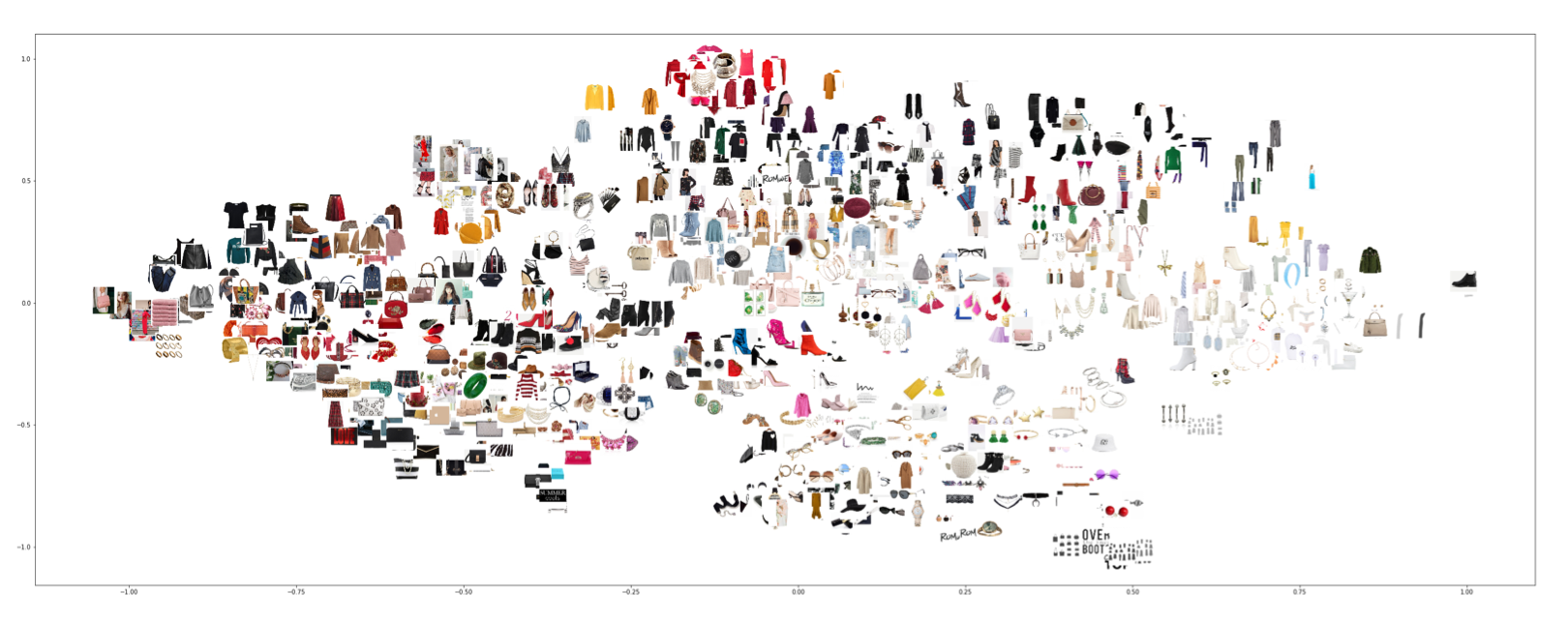}
    \label{fig:visualization_densenet}
    }
    \subfigure[FashionRN]{
        \includegraphics[width=.9\linewidth]{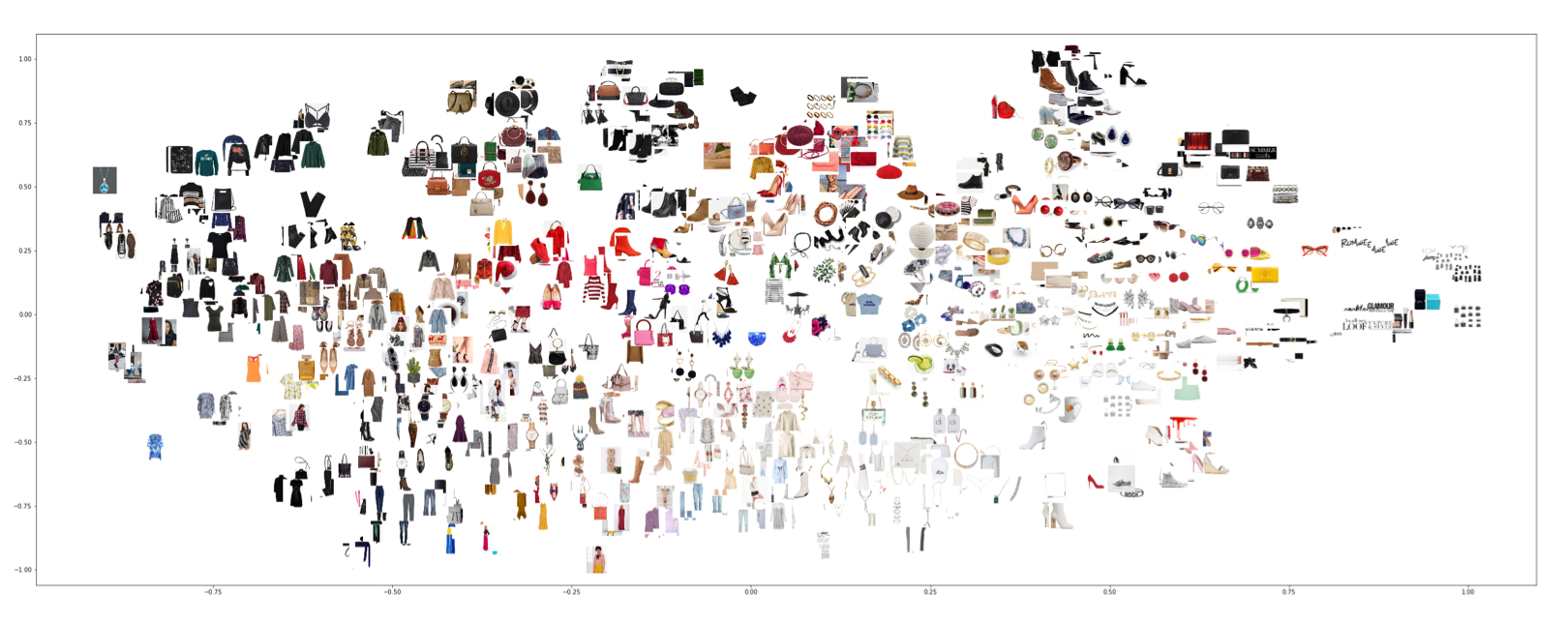}
    \label{fig:visualization_fashionrn}
    }
    \caption{Visualization of the same 1000 fashion items with embedding learned by different models.}
    \label{fig:visualization}
\end{figure*}

\subsection{Fill In The Blank (FITB) Test}
\label{sec:compatibility prediction}

In this task an outfit and a number of candidate items are given and the goal is to find the item that best matches with the existing items in the outfit. This is useful when a user has a set of items (e.g., a shirt and a pair of shoes) and wishes to find another item (e.g., a handbag) that best matches with the rest of the outfit. To run this test, we created a FITB dataset using our positive test set from section \ref{sec:Dataset}. In each test outfit, we randomly held-out one item. We then randomly selected three items from other outfits that have the same category as the held-out item. For example, if the held-out item was a shirt, all there randomly selected items were shirts. This is to ensure that the network cannot easily filter out items that already exists in the outfit without needing to understand compatibility. We then found the item among the four candidates that maximizes the compatibility score of the entire outfit.

Table \ref{table2} shows the results of FITB test for all the comparing methods.
Similar to the compatibility prediction task, we observed that our model outperforms all the baselines. The performance of this task is also improved through utilizing joint visual-semantic embeddings in our model (Fashion RN + VSE). 
The reason for this improvement in the FITB test is that in many cases there is more than one compatible item among the candidates. While the Fashion RN model is able to rely on visual information to find items that are compatible with the given outfit, adding semantic information can improve ranking among the compatible candidates. For example, in the last row of Figure \ref{fitb} the held-out item is correctly detected compatible (score = 0.98) by the Fashion RN model. However, the first shirt is also compatible with the outfit and has received a higher score. We observed that adding the semantic information (Fashion RN + VSE) improved ranking of the compatible candidates in this example and resulted in choosing the right shirt in the last row of Figure \ref{fitb}.

Similar to the compatibility prediction task, we observe that Bi-LSTM method performs poorly on our dataset. As others have noted \cite{hsiao2017creating} this is probably because the FITB test set provided by Han et al. \cite{han2017learning} contains poor choices of negatives. In their FITB dataset, negative items may not be the same type as the held-out item. For example, the test outfit may have a shirt, a pair of jeans, and be missing a handbag. If some of the candidates are shirts, which already exists in the outfit, the network can easily eliminate them based on their category without needing to infer compatibility.  Thus, to enforce the model to reason based on compatibility, we ensured that all the candidates are the same type as the missing item. Figure \ref{fitb} shows successful and unsuccessful examples of this test using our model.  In most of the test outfits the held-out item is among the top two items in the ranked list and shows a high compatibility score.

\section{Fashion Compatibility Embedding}
\label{sec:qualitative}

Besides the capability of predicting compatibility given a complete fashion outfit and fill-in-the-blank given an incomplete outfit, FashionRN and FashionRN-VSE are also able to learn item and outfit embedding through hidden layers. More specifically, $v$ learned in FashionRN can be viewed as the items' \emph{compatibility features}. As discussed previously, the concept of compatibility is fundamentally different from similarity, since items that are visually similar to each other are not necessarily compatible in fashion outfits, and vice versa.

To demonstrate the learned compatibility embedding of fashion items, we take the learned embedding, transformed them into two-dimensional embedding by using TSNE algorithm \cite{maaten2008visualizing}. To show FashionRN's capability of learning compatibility beyond visual similarity, we compare the visualization on the same set of randomly chosen 1000 fashion items with DenseNet features. The results are shown in Figure \ref{fig:visualization}.

As shown in Figure \ref{fig:visualization}, at the first glance, the scatters of the same 1000 fashion items created by DenseNet embedding and FashionRN embedding are greatly different. With a closer look, one can see that items with similar colors and shapes are closer to each other in the DenseNet embedding space, while items that make sense to go together in an outfit are closer to each other in the FashionRN embedding space. This shows that FashionRN captures the underlying item compatibility in addition to visual similarity.

\section{Conclusion}
\label{sec:conclusion}

In this paper, we proposed a method for learning fashion compatibility. We considered an outfit as a scene and its items as objects in the scene and
developed FashionRN and FashionRN-VSE, RN-based models, to learn the visual relations between items to determine their compatibility.
We collected a large dataset from Polyvore and conducted different experiments to demonstrate the effectiveness of our method. In addition to addressing some of the limitations of existing models, our model showed state-of-the-art performance in both compatibility prediction task and fill-in-the-blank test. Besides the capability in the above prediction tasks, FashionRN and FashionRN-VSE are also able to learn item and outfit embedding that carry underlying compatibility. To showcase such results, we visualize the learned embedding of the same items using both DenseNet and FashionRN. Through visualization, we find that FashionRN better capture the compatibility among items compared to DenseNet.



. 



\bibliographystyle{ACM-Reference-Format}
\bibliography{ref}
\end{document}